# The Use of Machine Learning Algorithms in Recommender Systems: A Systematic Review

**Ivens Portugal**
David R. Cheriton School
of Computer Science
University of Waterloo
Waterloo, ON, Canada
iportugal@uwaterloo.ca

**Paulo Alencar**
David R. Cheriton School
of Computer Science
University of Waterloo
Waterloo, ON, Canada
palencar@cs.uwaterloo.ca

**Donald Cowan**
David R. Cheriton School
of Computer Science
University of Waterloo
Waterloo, ON, Canada
dcowan@csg.uwaterloo.ca

## Abstract

Recommender systems use algorithms to provide users with product or service recommendations. Recently, these systems have been using machine learning algorithms from the field of artificial intelligence. However, choosing a suitable machine learning algorithm for a recommender system is difficult because of the number of algorithms described in the literature. Researchers and practitioners developing recommender systems are left with little information about the current approaches in algorithm usage. Moreover, the development of a recommender system using a machine learning algorithm often has problems and open questions that must be evaluated, so software engineers know where to focus research efforts. This paper presents a systematic review of the literature that analyzes the use of machine learning algorithms in recommender systems and identifies research opportunities for software engineering research. The study concludes that Bayesian and decision tree algorithms are widely used in recommender systems because of their relative simplicity, and that requirement and design phases of recommender system development appear to offer opportunities for further research.

## 1 Introduction

Recommender systems (RS) are used to help users find new items or services, such as books, music, transportation or even people, based on information about the user, or the recommended item [2]. These systems also play an important role in decision-making, helping users to maximize profits [15] or minimize risks [11]. Today, RSs are used in many information-based companies such as Google [35], Twitter [4], LinkedIn [50], and

Netflix [59]. The field of RS has its origins in the mid-1990s with the introduction of Tapestry [23], the first RS. As the RS field evolved, researchers studied the use of algorithms from machine learning (ML), an area of artificial intelligence (AI). Machine learning has been studied since the late 1950s [40], with the emergence of the field of AI. Today, there is a plethora of ML algorithms (k-nearest neighbor [48], clustering [28], Bayes network [22], to name a few types), which are used in applications that range from vacuum cleaner robots [13] and assistance for disabled people [30] to pattern recognition in images [63], or self-driving vehicles [62]. The potential application of ML algorithms is vast and the field looks very promising.

As previously mentioned, ML algorithms are being used in RSs to provide users with better recommendations. However, the ML field does not have a clear classification scheme for its algorithms, mainly because of the number of approaches and the variations proposed in the literature [37]. As a consequence, it becomes difficult and confusing to choose an ML algorithm that fits one's need when developing an RS. In addition, researchers may find it challenging to track the use of ML algorithms in RSs.

Software engineering (SE) is a field focused on the design, development and maintenance of software [49] with tools for organization, structure, and use of computer science knowledge [27]. RS and ML fields can benefit from the application of SE principles and definitions. In the area of RSs using ML algorithms, researchers may not know which SE areas could have significant impact.

One way to assist researchers and practitioners in choosing which ML algorithm to use in an RS, and to identify SE areas that can be improved in the development of RSs, is the study of the RS and ML fields. Research about RSs containing ML algorithms implemented in the literature can help show trends and provide a direction for future studies.

For these reasons, it was decided to do a systematic review to investigate how real implemented RSs with ML algorithms are being used; which ML algorithm they use; and how the SE field can impact their development. It is expected that, with this systematic review, researchers and practitioners can obtain more information about the RS field, and make better-informed implementation or research decisions.

This paper is organized as follows: section 2 describes the theoretical knowledge needed; Section 3 explains the systematic review plan, and section 4 explains its results. Section 5 finishes this study with conclusions and future work.

## 2 Theoretical Background

In this section, we give a brief overview of the fields of recommender systems and machine learning. This section focuses on a brief historical description with some definitions and classifications.

### 2.1 Recommender System

Recommender systems (RS) use artificial intelligence (AI) methods to provide users with item recommendations. For example, an online bookshop may use a machine learning (ML) algorithm to classify books by genre and then recommend other books to a user buying a specific book. RSs were introduced in 1992 when Tapestry, the first RS, appeared. Its authors used the term 'collaborative filtering' to refer to the recommendation activity. This term is still used to classify RSs. Recommender systems are divided into three main categories, depending on the information used to drive the recommendations: collaborative, content-based, and hybrid filtering [2].

RSs using a collaborative approach consider the user data when processing information for recommendation. For instance, by accessing user profiles in an online music store, the RS has access to all the user data, such as the age, country, city, and songs purchased. With this information, the system can identify users that share the same music preference, and then suggest songs bought by similar users.

RSs with a content-based filtering approach base their recommendations on the item data they can access. As an example, consider a user who is looking for a new computer using an online store. When the user browses a particular computer (item), the RS gathers information about that computer and searchers in a database for computers that have similar attributes, such as price, CPU speed, and memory capacity. The result of this search is then returned to the user as recommendations.

The third classification describes RSs that combine the two previous classifications into a hybrid filtering approach, recommending items based on the user and the item data. For example, on a social network, an RS may recommend profiles that are similar to the user (collaborative filtering), by comparing their interests. In a second step, the system may consider the recommended profiles as items and thus access their data to search for new similar profiles (content-based filtering). In the end, both sets of profiles are returned as recommendations.

When using a collaborative or a hybrid filtering approach, RSs must gather information about the user in order to develop recommendations. This activity can be done explicitly or implicitly. Explicit user data gathering happens when users are aware they are providing their information. For instance, when registering for a new online service, users usually fill in a form that asks their name, age, and email. Other forms of explicit user data gathering are when users express their preferences by rating items using a numerical value or a preference such as a Facebook "like." Implicit user data gathering accesses information about the user indirectly. For example, when visiting an online store, the server at the online store exchanges messages with the user's computer, and based on that, the store's RS may know the browser the user is using, as well as the user's country. More advanced applications monitor user click and keystroke logs.

Besides the common recommendation process, in which users are presented with items that might be of interest, recommendations can be provided in other ways. Trust- based recommendations [45] take into consideration the trust relationship that users have between them. A trust relationship is a link in a social network to a friend or a following connection. Recommendations based on trusted friends are worth more than those that do not have trust links. Context-aware recommendations [3] are based on the context of the user. A context is a set of information about the current state of the user, such as the time at the user location (morning, afternoon, evening), or their activity (idle, running, sleeping). The amount of context information to be processed is high, making context-aware recommendations a challenging research field. Risk-aware recommendations [11] are a subset of context-aware recommendations and take into consideration a context in which critical information is available, such as user vital signs. It is risk-aware because a wrong decision may threaten a user's life or cause damage. Some examples are recommending pills to be taken or stocks the user should buy or, sell.

## 2.2  Machine Learning

Machine Learning (ML) uses computers to simulate human learning and allows computers to identify and acquire knowledge from the real world, and improve performance on some tasks based on this new knowledge. More formally, [43] defines ML as follows: "A computer program is said to learn from experience E with respect to some class of tasks T and performance measure P, if its performance at tasks in T, as measured by P, improves with experience E". Although the first concepts of ML originated in the 1950s, ML was studied as a separate field in the 1990s [43]. Today, ML algorithms are used in several areas besides computer science, including business [6], advertising [16] and medicine [32].

Learning is the process of knowledge acquisition. Humans naturally learn from experience because of their ability to reason. In contrast, computers do not learn by reasoning, but learn with algorithms. Today, there are a large number of ML algorithms proposed in the literature. They can be classified based on the approach used for the learning process. There are four main classifications: supervised, unsupervised, semi-supervised, and reinforcement learning.

Supervised learning happens when algorithms are provided with training data and correct answers. The task of the ML algorithm is to learn based on the training data, and to apply the knowledge that was gained in real data. As an example consider an ML learning algorithm being used in book classification in a bookstore. A training set (training data + answers) can be a table relating information about each book to a correct classification. Here, information about each book may be title, author, or even every word a book contains. The ML algorithm learns with the training set. When a new book arrives at the bookstore, the algorithm can classify it based on the knowledge about book classification it has acquired.

In unsupervised learning, ML algorithms do not have a training set. They are presented with some data about the real world and have to learn from that data on their own. Unsupervised learning algorithms are mostly focused on finding hidden patterns in data. For example, suppose that an ML algorithm has access to user profile information in a social network. By using an unsupervised learning approach, the algorithm can separate users into personality categories, such as outgoing and reserved, allowing the social network company to target advertising more directly at specific groups of users.

ML algorithms can also be classified as semi-supervised. Semi-supervised learning occurs when algorithms work with a training set with missing information, and still need to learn from it. An example is when an ML algorithm is provided with movie ratings. Not every user rated every movie and so there is some missing information. Semi-supervised learning algorithms are able to learn and draw conclusions even with incomplete data.

Lastly, ML algorithms might have a reinforcement learning approach. Reinforcement learning occurs when algorithms learn based on external feedback given either by a thinking entity, or the environment. This approach is analogous to teaching dogs to sit or jump. When the dog performs the action correctly, the dog receives a small cookie (positive feedback). It does not receive any cookie (negative feedback) if it performs the wrong action. As an example in the computer science field, consider an ML algorithm that plays games against an opponent. Moves that lead to victories (positive feedback) in the game should be learned and repeated, whereas moves that lead to losses (negative feedback) are avoided.

ML has become quite popular recently with the increase in processor speed and memory size. As a consequence, the field now has a large number of algorithms that use mathematical or statistical analysis to learn, draw conclusions or infer data. This number continues to increase as evidenced by the number of scientific publications that propose variations or combinations of ML algorithms. For that reason, ML algorithms have been categorized based on the purpose for which they are designed. Some examples of classification can be found in [56], and [34], although the field still does not have any standards.

# 3   Systematic Review

When developing RSs, software engineers must decide on the specific recommender algorithm of all those available. This choice has significant effect on the rationale of the RS, on the data that will be needed from users and recommendation items, and on performance issues. The number of algorithm variations and combinations in the literature makes this choice a challenging task.

This large number of recommender algorithms, which appears to be constantly growing and changing, makes software engineering for RSs a continuing challenge. Trying to develop tools to make RS development easier is a moving target, as new studies must be done to observe new open problems and trends, and further enrich the knowledge base.

For these reasons the authors decided to do a systematic review of the ML field to analyze the development of RS containing ML algorithms and to see if there could be a more organized approach to software engineering of RSs. This review has two main goals:

1. to identify which ML algorithms are used in RSs and
2. to discover open questions in RS development that might be impacted by SE.

This study has one restriction. The authors decided to limit the set of scientific publications to those describing a case study, or implementation. A case study is a scientific method that describes real world phenomena, either by observing or applying a new technology. In the scope of this work, the authors are concerned with finding publications that were implemented and validated with real data. The restriction is formally defined as follows:

- R1: The study investigates RS approaches that contain ML algorithms and that were implemented and tested with real or simulation data.

The main reason for this restriction is that several publications in the literature propose new algorithms that are never tested or validated.

This review examines the following two research questions (RQ):

- RQ1: What are the most used machine learning algorithms in recommender systems?
- RQ2: What are the software engineering areas that can help answer open questions in the development of recommender systems containing machine learning algorithms?

To answer the first question, the authors read each publication returned by a search query and list the proposed algorithms. The approach to the second question is different. First, the authors limit the scope of SE areas to the five stages of the waterfall model for software lifecycle: requirements, design, implementation, verification, and maintenance [49]. The waterfall model describes the high-level stages that any software development project should follow. Next, the authors read the conclusion and future work sections of every publication and identifies problems or research opportunities that need to be investigated. Finally, the authors classify the problem or research opportunity into one of the five software lifecycle stages. Through this classification, the authors expect to have a perspective on which areas of SE research would be valuable in supporting the implementation of RSs.

To strengthen the validity of the review the authors applied certain exclusion criteria (EC) to the publications that were included. These criteria and the rationale are presented next.

- EC1: Publications must have been peer-reviewed and published in a conference, journal, press, etc. Publications that were not peer-reviewed or formally published are excluded.
- EC2: Publications must describe the recommender system and the machine learning algorithm sufficiently well that it is possible to identify the algorithm. Publications that do not clearly state the machine learning algorithm being used are excluded.
- EC3: Publications must contain the description of a case study or the implementation of the approach that is proposed. Publications that do not describe a real life implementation are excluded.
- EC4: Publications must be primarily in English. Publications in languages other than English are excluded.
- EC5: Books, letters, notes, and patents are excluded.
- EC6: Publications must be unique. If a publication is repeated, other copies of that publication are excluded.

There are several synonyms for RSs. Based on [29] this systematic review considers RS terms that replace "recommender" by "recommendation", and terms that replace "system" by "platform", or "engine". This systematic review does not consider any "machine learning" synonyms.

The following search query (RQ) was created for this systematic review:

- RQ: TITLE-ABS-KEY(((recommender OR recommendation) PRE/0 (system OR engine OR platform)) AND "machine learning" AND (implementation OR "case study"))

This search query inspects publication title, abstract and keywords and attempts to find terms that relate to the field of RS, ML, and provide some indication that the proposed approach was implemented. PRE/0 is a search engine-specific syntax to express the relationship between two terms. In this search query, PRE\0 restricts the results to publications that have "recommender" (or its synonyms) followed by "system" (or its synonyms). Publications must also contain the term "machine learning" in the title, abstract, or keywords. To retrieve implemented approaches, the search query also looks for the terms "implementation" or "case study".

The search query was used on the popular academic search engine Scopus [54] on July 24, 2015. The search returned 35 publication entries that had to be reviewed for quality.

Some publication entries were excluded based on the exclusion criteria previously explained, and they are summarized on Table 3.1.

Table 3.1 - Publications that were included and excluded from the systematic review.

| Total retrieved | | 35 |
|---|---|---|
| **Reason** | **Publication** | **Total** |
| Conference / Proceedings entries | | 3 |
| Not describing a case study or implementation | [46] | 1 |
| Not sufficiently describing the approach | [38] [51] [52] | 3 |
| Not able to access | [47] | 1 |
| Other language | | 0 |
| Duplicated | [52] | 1 |
| Books, letters, notes, and patents | | 0 |
| Total retained | [5] [7] [8] [9] [10] [12] [14] [19] [20] [21] [24] [25] [31] [36] [39] [41] [53] [55] [57] [58] [60] [61] [64] [64] [67] [68] | 26 |

The number of publications to be read in the systematic review decreased from 35 to 26 when filtered by the exclusion criteria. Three of the publication entries were conference or proceeding descriptions and are excluded because they are not written scientific work. The authors did not have access to one publication, even after asking help for colleagues and visiting libraries. This publication then was excluded from the systematic review. Scopus seems to have a minor issue because it listed the same publication twice. However, one copy of this publication was excluded according to EC6. The remaining 30 publications were downloaded and assessed for quality. The authors read the abstract and the approach description and, if present, the case study description. Based on that, one publication was excluded for not describing a case study (EC3) and three others were excluded due to poor description of the approach and the ML algorithm (EC2). In the end, 26 publications were retained and ready to be analyzed.

## 4 Systematic Review Results

The reading process was focused on finding three pieces of information: the ML algorithm, the domain of the case study or implementation, and the problems or open questions associated with the proposed work. Therefore, after reading the abstract and the introduction, the authors read the approach and case study description, and finally the conclusion or future work of each publication included in the systematic review.

The authors developed a spreadsheet with an identification of each publication alongside three spaces for noting the pieces of information previously described. After reading all publications, the authors processed the information contained in the spreadsheet and organized it in a presentable manner. The results and conclusions are presented in the following paragraphs.

Processing ML algorithm names means that the authors separated algorithms into categories based on [56]. Some algorithms had obvious classifications because they were small variations of well-established algorithms (e.g. incremental matrix factorization is a variant of the matrix factorization algorithm). However, some algorithms do not intuitively fit in any category. For these cases, the algorithm was listed under a new category with its own name (e.g. Topic Independent Scoring Algorithm is on category TISA). Table 4.1 shows the results.

Table 4.1 - Types of machine learning algorithms used in recommender systems.

| Category | Publications | Total |
|---|---|---|
| Bayesian | [20] [21] [24] [36] [39] [55] [66] | 7 |
| Decision Tree | [21] [36] [39] [55] [57] | 5 |
| Matrix factorization-based | [7] [53] [60] [61] | 4 |
| Neighbor-based | [25] [31] [39] [60] | 4 |
| Neural Network | [5] [21] [39] [41] | 4 |
| Rule Learning | [24] [25] [36] [58] | 4 |
| Ensemble | [10] [21] [57] | 3 |
| Gradient descent-based | [8] [14] [60] | 3 |
| Kernel methods | [14] [21] [66] | 3 |
| Clustering | [19] [20] | 2 |
| Associative classification | [36] | 1 |
| Bandit | [12] | 1 |
| Lazy learning | [21] | 1 |
| Regularization methods | [14] | 1 |
| Topic Independent Scoring Algorithm | [41] | 1 |

When inspecting Table 4.1, one may notice that Bayesian approaches are used as the algorithm for RSs in 7 out of the 26 publications read. The authors theorize that this result is because of the reduced complexity of Bayesian calculations. The same explanation may justify the use of Decision Trees as the second most used approach for ML algorithms in RS. Both Bayesian approaches and Decision Trees have similar underlying calculations and recently have become popular.

Neural networks are also popular and their usage seems promising, with applications in self-driving cars and image inspection and classification. Although, neural network approaches were found in four publications, much more work could be performed in using algorithms of this type for recommending items or services to users. One reason that explains a lower adoption rate for neural networks in RS field is that researchers still do not fully understand why some neural networks make a particular decision. Understanding neural networks is still a research topic.

Two other ML areas that became popular in the last two decades are Bandit algorithms and Ensemble approaches, but both had a low number of publications in this systematic review. Bandit algorithms are used to decide how many times and in which order decisions are made to maximize a given value [65]. The high complexity of the Bandit approach may be a significant factor in preventing the adoption of this algorithm in RSs. Ensemble learning [33] became increasingly popular especially after the Netflix Prize

[42] in 2009. The notion is to combine several results from different algorithms into a large knowledge base and draw conclusions from it. Similar to the Bandit algorithm, the complexity of implementing and generating results appears to impede its growth in the RS field.

While reading the 26 publications of this systematic review, the authors noted that some of them described the ML algorithm with some mathematical or statistical method. For example, [12] describes a multi-armed Bandit algorithm with cosine measure. Other works described the ML algorithm in a distributed environment with Map Reduce. Table 4.2 depicts the publications that used a mathematical or statistical method, and those who used the MapReduce algorithm.

Table 4.2 - Publications that described a machine learning algorithm using a mathematical or statistical method, and publications that did it with the MapReduce algorithm.

| Method | Publications | Total |
| --- | --- | --- |
| Mathematical / Statistical | [7] [8] [9] [10] [12] [14] [20] [25] [39] [53] [64] [67] [68] | 13 |
| MapReduce | [19] [20] [53] [61] | 4 |

Mathematical or statistical methods are calculations such as cosine measure [18], least squares [1] or Pearson correlation [18]. They were listed separately, because ML algorithms use them to generate parameters that help in recommendations. It is expected that the great majority of publications describe a mathematical or statistical method because the ML research field has its roots in these areas. The MapReduce algorithm [17], which became popular in mid-1990s, is a programming model suited for distributed and parallel computation. Four of the publications described their approach in an environment in which the MapReduce algorithm was used. Although it was not the main algorithm, MapReduce presence shows the likely importance of the field of Big Data in the future of RSs.

The authors, when reading the case study or implementation description, noted the domains in which the publications validate their proposals. The identification of these domains enriches the knowledge of the field and helps future researchers when validating their work. Table 4.3 shows the results. Note that some publications appear more than once, either in the same or in different rows. It happens because some publications reported more than one case study or implementation.

The domain of Movies is used the most with 10 occurrences in the 26 publications. One reason for this result is the ease of access to data in the movie domain. The University of Minnesota maintains a dataset with several movie ratings, named MovieLens [44], which is widely used in several RS case studies. Another source of user ratings in the movie domain is the Internet movie database (IMDb) [26], which contains millions of titles and ratings that can be gathered to build a testing dataset. The domains of Documents and Product reviews had significant results with four occurrences each. Text processing is one of the most popular forms of data gathering. Several algorithms are aimed at extracting information from text sources, and this is the main factor behind the position of both domain areas in this ranking.

Table 4.3 - Domains of case studies and implementations described in the publications.

| Domain | Publications | Total |
|---|---|---|
| Movie | [5] [5] [5] [14] [36] [39] [53] [53] [60] [60] | 10 |
| Documents | [9] [19] [20] [20] | 4 |
| Product review | [25] [51] [57] [58] | 4 |
| Jokes | [14] [60] | 2 |
| Music | [14] [53] | 2 |
| TV | [58] [68] | 2 |
| Academic | [61] | 1 |
| Books | [36] | 1 |
| Elections | [64] | 1 |
| E-mail | [24] | 1 |
| E-shop | [21] | 1 |
| Mobile navigation | [12] | 1 |
| Safety-critical process automation | [10] | 1 |
| Social | [31] | 1 |
| Stocks | [67] | 1 |
| Telecommunications | [10] | 1 |
| Tourism | [66] | 1 |
| Traffic | [55] | 1 |
| Webpages | [8] | 1 |

The second research questions (RQ2) is concerned with the software engineering areas where research effort can be improved in RS development with an ML algorithm. After listing five SE high-level areas, taken from the Waterfall model for software development, the authors read both the conclusion and the future work sections of publications looking for the problems and future work described. Later, the authors separated the results into the five areas. The final result is shown in Table 4.4.

Table 4.4 - Number of publications that reported problems in each of the five software engineering areas considered in this study.

| SE area | Publications | Total |
|---|---|---|
| Requirements | [5] | 1 |
| Design | [25] [24] [55] | 3 |
| Implementation | [10] [12] [19] [24] [25] [39] [61] [64] [66] | 9 |
| Verification | [8] [10] [12] [20] [24] [25] [39] [51] [53] [61] [64] [67] | 12 |
| Maintenance | | 0 |

Roughly 80% of the publications describe problems or future work that focus on the implementation or verification areas. This result highlights the importance of these areas in RS development. However, limiting the search query results to publications that have a case study or an implementation description had an impact on the number of problems and future work reported. In contrast, only four publications were concerned with problems or future work in requirements, design or maintenance of RSs. This implies that little effort is being spent on these areas and several open questions may exist.

## 5 Conclusion & Future Work

Currently, recommender systems (RS) are widely used in e-commerce, social networks, and several other domains. Since its introduction in mid 1990s, research in RSs has been evolving. One progressive step in RS history is the adoption of machine learning (ML) algorithms, which allow computers to learn based on user information and to personalize recommendations further. Machine learning is an Artificial Intelligence (AI) research field that encompasses algorithms whose goal is to predict the outcome of data processing. ML has made major breakthroughs in the fields of image recognition, search engines, and security. However, the ML field has several algorithms described in the literature, with varied characteristics. The literature lacks a classification system for algorithms showing the environment in which they are most suitable. Therefore, choosing an ML algorithm to be used in RSs is a difficult task. In addition, researchers in RSs do not have a clear view of the trends in ML algorithm usage to decide on where to focus their research efforts. Software engineering (SE) is a field that studies the development of computer software from concept to implementation, and maintenance. The field has tools that can assist the development of RSs that contain ML algorithms. However, researchers need to know in which SE area RS development lacks resources. This study then proposes a systematic review to observe the ML algorithms that are used in RSs and what SE areas can assist the development of such RSs.

The systematic review collected 26 publications, after filtering out some based on exclusion criteria. All publications were read and the conclusions are as follows. In RS development, the ML algorithm used is a Bayesian or decision tree approach in most of the cases. Their recent popularity and the low complexity in calculation and implementation contribute to this result. Moreover, most of the approaches have a mathematical or statistical description of algorithms, since the ML field has origins in mathematics and statistics. Four of the publications have proposed RSs that work with the MapReduce algorithm. This result shows that the field of Big Data is relevant in RS studies, allowing even more personalized recommendations.

During the systematic review, the authors investigated the domains in which proposals of RSs with an ML algorithm were validated. The three most used domains are movies, documents, and product reviews, mainly because of the ease in accessing test data. For the movie domain, MovieLens and IMDb are two online datasets of movie ratings. For the documents and product review domains, the authors note that text is one of the most common forms of processing data, and several ML algorithms were created for text processing.

The second goal of this study concerns analyzing the development of RSs that have an ML algorithm and discovering problems or open questions in which SE areas can assist. The problems or open questions were found by reading the conclusion and future work sections of publications, and they were classified in one of the five areas of the Waterfall model of the software lifecycle. The results show that researchers report more problems and open questions in implementation and verification areas. However, as the search question looks for case study or implementation description in the publications, this affected the results. The lack of problems or open questions reported in the other SE areas (requirements, design, and maintenance) may indicate that there are research opportunities to be investigated.

This study serves as a basis for investigating RS development. In the future, more studies on the use of Bayesian algorithms in RSs can be done to observe the implications of their use, performance, and utility. Moreover, RS development lacks studies analyzing early stages, such as requirements and design, and late stages, such as maintenance. Open questions in these stages must be investigated to improve the knowledge about the field.

# 6 Acknowledgments

The authors would like to thank the Natural Sciences and Engineering Research Council of Canada (NSERC), the Ontario Research Fund of the Ontario Ministry of Research and Innovation, SAP, and the Centre for Community Mapping (COMAP) for their financial support to this research.